# ELECTRONIC PROPERTIES OF GROUP III-A NITRIDE SHEETS BY MOLECULAR SIMULATION

[1]E. Chigo Anota, [2]M. Salazar Villanueva, [1]H. Hernández Cocoletzi

[1]Cuerpo Académico de Ingeniería en Materiales,
[1]Facultad de Ingeniería Química, [2]Facultad de Ingeniería
Avenida San Claudio y 18 sur s/n CU, Edificio 106A, Puebla
72570, México
e-mail: [1]echigoa@yahoo.es

We have performed first principles total energy calculations to investigate the structural and reactivity parameters of novel $N_{12}X_{12}H_{12}$ (X=B, Al, Ga, In, Tl) nitrides, in their coronene-like ($C_{24}H_{12}$) structure to simulate these sheets. The exchange and correlations potential energies are treated in the generalized gradient approximation (GGA), and the local density approximation within the parameterization of Perdew-Wang and Perdew-Burke-Ernzerhof (PWC, PBE) and the doubly polarized atomic base (DNP). The chemical potential, hardness and electrophilicy index, as well as bond length are reported. The bond length of the structures is similar to the bulk. The gap between the HOMO and LUMO decreases from BN (5.18eV) to TlN (1.76 eV). At the same time, the polarity increases except for TlN.

**Key words**: coronene, nitrides, DFT theory.

**PACS: 31.10+z, 31.15.Ar, 31.15Ew**

## 1. BACKGROUND

The discovery and synthesis of graphene has motivated intense investigations on this kind of nanostructures. The first of this system was the boron nitride (BN) sheet [1]. Both graphene and BN sheets, are attractive due their important applications, as gas sensors [2], in the possible increase on the information storage in the DVD´s [3] and nanoelectronic applications [4]. Recently, Yge [5] has studied the phase stability of a new graphene like structure, the boron carbon nitride ternary alloy, opening the possibility of usage of these systems. In particular, systems containing nitrogen have been very attractive because their fabulous applications.

Here, we report first principles total energy calculations to investigate the physicochemical properties of novel $N_{12}X_{12}H_{12}$ (X=B, Al, Ga, In, Tl) nitrides, in their coronene ($C_{24}H_{12}$) like structure. This kind of two dimensional systems has shown capability to adsorb atomic [6] and molecular hydrogen [7]. Additionally, it has been shown the adsorption of water molecule by the circular boron nitride sheet [8] and doped with carbon [9]. Finally, the stability of the Li, Na, F, and Cl doped BN sheet has been studied, showing that the methodology we used is appropriate [10]. The calculations are done within the density functional theory (DFT), using the local density approximation (LDA) and the generalized gradient approximation (GGA), within the parameterizations developed by Perdew-Wang [11] and Perdew-Burke-Ernzerhof [12]. The optimum structural parameters as well as dipolar moment, vibration frequencies, chemical potential, hardness and electrophilicy index, are calculated.

## 2. COMPUTATIONAL PROCEDURE

We performed first principles total energy calculations to study the atomic configuration of the coronene like novel $N_{12}X_{12}H_{12}$ (X=B, Al, Ga, In, Tl) nitrides. Calculations are done within the DFT formalism as developed in the DMOL$^3$ code, available from Accelrys Inc. [13]. The exchange and correlation potential energies are treated within the generalized gradient approximation (GGA), and the local density approximation (LDA) within the Perdew-Wang (PWC) and Perdew-Burke-Ernzerhof (PBE) parameterization. The doubly polarized all electron atomic base (DNP) for the core, in the singlet ground state (charge = 0, multiplicity = 1), was used. The limit for BN, AlN, GaN, and InN orbital was of 0.4 nm and 0.5 nm for TlN; the convergence for the SCF cycles was $1.0 \times 10^{-6}$ Ha. The obtainment of non-negative frequencies was the criterion for structural stability [14].

The clusters contain 36 atoms (12 Nitrogen atoms; 12 atoms of B, Al, Ga, In, and Ti; 12 hydrogen atoms), Figure 1. The reactivity parameters are obtained in terms of the HOMO and LUMO; the hardness ($\eta$) is the difference between HOMO and LUMO, the chemical potential is obtained through $\mu=(LUMO+HOMO)/2$ [15], and the electrophilicy index with $\varepsilon=\eta^2/2\mu$ [16].

## 3. RESULTS AND DISCUSSION.

The geometrical optimization was done under room temperature and 1 atm. Table 1 presents results on the structural parameters using GGA and LDA approximations; as we can see, there is no significant difference between both approximations, below the 2.5 % in all cases, indicating that exchange-correlation effects are not important. The bond length of the clusters increases as the atomic number increases and the hexagons are almost regular; the lattice parameter of the bulk is also included. The bond angle N-X-N for BN, AlN, and GaN is $118.79^0$ and $119.45^0$ for InN and TlN, very near from the ideal value. In all cases the vibration modes were positive, ensuring the structural stability [14].

Following the analysis, we observe that the biggest value for the heat capacity (pressure constant), within LDA approach, corresponds to TlN, which is bigger than graphite (2.04 KJ/mol K) and water (18.0 KJ/mol K a 25 $^0$C), suggesting an application as a ceramic at the nanoscale. On the other hand, the entropy increases from BN to TlN about 32 %. i. e. the clusters are more disordered as the atomic number increases. The enthalpy decreases from BN to TlN about 23 %; this means that the capacity for exchange heat of the systems decreases. The same conclusions are obtained for the GGA approach.

On the other hand, the dipole moment increases as the atomic number increases, from $6.4 \times 10^{-3}$ Debye for BN to $3.8 \times 10^{-1}$ for TlN (except for GaN) in the LDA approach, and from $7.1 \times 10^{-2}$ for BN to $3.47 \times 10^{-1}$ Debye for InN, in the GGA approach; it means that there is a tendency to increase the covalence as we go on to the TlN.

Table 2 contains results on the chemical potential (μ), electrophilicy index (ε), and the hardness (η). In the case of LDA approach, the lower value of μ corresponds to GaN and AlN, and there is a little increase for BN; the biggest value is for InN and TlN (about 12%). However, for GGA, BN, AlN and GaN, have almost the same value, while InN presents the biggest one. The electrophylici index follows a monotone decrease (90%) from BN to TlN, in both, LDA and GGA approaches. This result indicates that the thallium nitride redistributes its charge more easily in the presence of a charge, and in a possible adsorption process it would be a physisorption like.

In Table 2 is also shown the gap (hardness is this work) between the energy of the highest occupied molecular orbital (HOMO) and the energy of the lowest unoccupied orbital (LUMO), p in character. The highest value corresponds to BN (5.18), similar to the bulk (6.20 eV). There is a decrease of this gap from BN to TlN (1.76); in the nomenclature of semiconductors, these values correspond to an insulator for BN to a semiconductor for TlN [9]. Something similar happens in the bulk for wurtzite and zinc blende structures. According to the definition of Pearson and Parr [17], on the reactivity, this parameter indicates the facility for a redistribution of the charge in the system; the value obtained for TlN indicates a relative facility to redistribute its charge, as concluded with the electrophilic index. Finally, the molecular orbitals are concentrated around the central hexagon of the structure, i. e. the electronegative zone is in this region; a previous work confirm this asseveration [9]. The later discussion is valid for both, LDA and GGA approximations.

## 4. CONCLUSIONS

We have used a molecular approach to investigate the structural and the reactivity parameters of the coronene like nitrides, at room temperature and 1 atm. The study was done using first principles calculations within the GGA and LDA approximations for the correlation-exchange termn. The bond length is almost the same for both, GGA and LDA, approximations, i. e., correlation-exchange effect is not important for this parameter. The covalence increases from BN to TlN, and the thermodynamic properties suggest applications as a ceramic at the nanoscale, mainly the TlN. The gap between the HOMO and the LUMO is similar to the value of the bulk. Finally, we have shown that this methodology is appropriate to study this kind of nanostructures and hence, to use it to investigate others systems.

**Acknowledgements**

The authors acknowledge the support to this work of: VIEP-BUAP (No. CHAE-ING08-I), Proyecto Interno de Investigación-FIQ-BUAP (2008-2009) and Proyecto de Ciencia Básica-CONACYT, México (No. 0083982).

# Figure captions

**Figure 1**. Coronene-like sheet nitrides ($N_{12}X_{12}H_{12}$; N: in blue; X= (B, Al, Ga, In, Tl): in pink and in white the hydrogen)

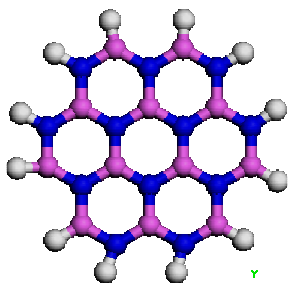

**FIGURE 1**

**Table 1.** Optimized geometric parameters, dipole moments and thermodynamic properties (entropy, heat capacity at constant pressure and enthalpy).

| Functional/base | Length X-N (Å) | Error % Exp. Vs Theor. | Dipole Moment (Debye) | S      Cp      H Kcal/mol |
|---|---|---|---|---|
| **Experimental** [18] | 1.421 (BN) | | | |
| **LDA(PWC)/DNP** Sheet | | | | |
| BN | 1.44 [19] | 1.34 | 0.0064 | 5849 / 10.23 / 158.59 |
| AlN | 1.79 | | 0.096 | 66.62 / 15.94 / 111.43 |
| GaN | 1.84 | | 0.205 | 72.91/ 19.15/ 105.03 |
| InN | 2.06 | | 0.033 | 84.36/ 28.84/ 92.556 |
| TlN | 2.15 | | 0.381 | 98.47/ 39.84 / 85.52 |
| **GGA(PBE)/DNP** Sheet | | | | |
| BN | 1.455 | 2.39 | 0.071 | |
| AlN | 1.82 | | 0.124 | |
| GaN | 1.87 | | 0.259 | |
| InN | 2.097 | | 0.347 | |
| TlN | 2.20 | | | |

[18] Pauling, Proc. Natl. Acad. Sci. **56,** 1646 (1966).
[19] R. G. Parr, L. V. Szentpály, S. Liu, J. Am. Chem. Soc. **121**, 1922 (1999).

**Table 2.** Chemical reactivity parameters for the stable configuration (in absolute value).

| System | Chemical potential (μ) (eV) | Electrofily (ε) (eV) | Hardness chemical/Gap energy (eV) |
|---|---|---|---|
| **LDA(PWC)/DNP Sheet** | | | |
| BN | 3.74 | 3.57 | 5.18 |
| AlN | 3.62 | 2.91 | 4.59 |
| GaN | 3.67 | 2.11 | 3.94 |
| InN | 4.195 | 0.65 | 2.33 |
| TlN | 4.25 | 0.36 | 1.76 |
| **GGA(PBE)/DNP Sheet** | | | |
| BN | 3.50 | 3.50 | 5.17 |
| AlN | 3.42 | 3.42 | 4.51 |
| GaN | 3.54 | 3.54 | 3.66 |
| InN | 4.02 | 4.02 | 2.11 |
| TlN | - | - | - |
| **Bulk wurtzite** | | | |
| BN | | | - |
| AlN | | | 6.20 |
| GaN | | | 3.44 |
| InN | | | 0.75 |
| TlN | | | - |
| **Zinc-blende** | | | |
| BN | | | - |
| AlN | | | 5.11 (Theory) |
| GaN | | | 3.2-3.3 |
| InN | | | 2.2 (Theory) |
| TlN | | | - |